# Multistep severe plastic deformation to achieve non-rare earth bulk magnets with high α-MnBi phase content

L. Weissitsch[a], S. Wurster[a], H. Krenn[b], A. Bachmaier[a]*

[a] *Erich Schmid Institute of Materials Science, Austrian Academy of Sciences, Jahn Strasse 12, 8700 Leoben, Austria;* [b] *Institute of Physics, University of Graz, Universität Platz 5, 8010 Graz, Austria, *Corresponding author: andrea.bachmaier@oeaw.ac.at*

# Multistep severe plastic deformation to achieve non-rare earth bulk magnets with high α-MnBi phase content

The ferromagnetic α-MnBi phase as non-rare earth magnetic material has gained increasing interest, but fabrication of large volumes containing a significant amount of α-MnBi is still challenging. Targeting successful processing strategies, we apply multistep severe plastic deformation with intermediate magnetic field assisted annealing. The subsequent severe plastic deformation renewed the high amount of material defects and lead to microstructural refinement of the previously formed α-MnBi phase. Thus, α-MnBi phase content is enhanced during final annealing. Secondly, the magnetic coercivity increases. These results suggest that further optimization will pave the way towards non-rare-earth bulk magnetic materials with enhanced α-MnBi phase content.

**Impact statement**

A novel solid-state processing route consisting of multistep severe plastic deformation and magnetic field annealing is presented, which opens a new pathway for processing rare-earth free bulk magnetic materials.

**Introduction**

As current high-performance permanent magnets contain a significant amount of rare-earth elements, an important goal is to find a rare-earth free alternative [1]. One candidate material is α-MnBi, which is an intermetallic low temperature phase with exceptional hard magnetic properties. It has a positive temperature coefficient of intrinsic magnetic coercivity for T<500 K, so its intrinsic coercive force increases with rising temperatures [2,3]. Thus, a large and peaking magnetocrystalline anisotropy of about 2.2 MJ/m$^3$ above 400 K is reached, which makes this material interesting to be used as permanent magnet for elevated temperature applications [4,5].

The challenging problem lies in the difficult fabrication process of ferromagnetic single phase α-MnBi. According to the phase diagram, α-MnBi is a line compound with equiatomic composition. This phase exists below a temperature of about 355°C, but decomposes due to eutectic transition above 262°C [6,7]. Due to its limited low temperature stability, but also because of large variety of different Mn-Bi phases above the mentioned temperatures, a direct fabrication of the α-MnBi phase with classical metallurgical methods is restricted to only limited cases [8].

High volume fraction of the α-MnBi phase is obtained, for example, by a sequence of different processing steps often including non-equilibrium conditions [9–17]. For example, in[9] extremely rapid cooling of a Mn-Bi melt is applied, followed by a milling process to produce a fine powder. This powder is magnetically sieved and then compressed into a pellet. The choice of chemical composition associated with annealing treatments allows the formation of an intermediate phase (β-MnBi phase) and a subsequent transformation to the α-MnBi phase.

The complexity of the necessary processing steps is challenging. Each step might be subject to a certain variance of the process parameters. Furthermore, the pronounced tendency to oxidation of the starting material Mn, but also of the resulting α-MnBi phase must be taken into consideration [18,19]. Oxidation during the process steps inevitably leads to a loss of available Mn, resulting in a reduction of the partition of the α-MnBi phase, which necessitates an inert atmosphere during processing, especially in grinding and powdering steps. For industrial utilization a simplified processing route might be beneficial [1] and this is why we propose the simple concatenation of the sequence "deformation-annealing" as a prospective way for an improved α-MnBi phase formation.

In a recent work, bulk Mn-Bi composite material was produced from an elemental Mn and Bi powder mixture by severe plastic deformation (SPD), resulting in grain and phase refinement as well as shear deformation induced magnetic anisotropy [20]. The high crystallographic defect density introduced by the SPD process is crucial for accelerated diffusion processes during the subsequent thermal treatment. An additional applied magnetic field during annealing had a positive effect on the α-MnBi phase formation. We thus aim to apply a second SPD step on this prepared bulk Mn-Bi composite material to study the influence of an additional deformation step on the α-MnBi phase formation during subsequent magnetic field annealing. By analysing the microstructure and magnetic properties, we find that the second SPD step improves homogeneity of the Mn-Bi composites and leads to microstructural refinement of the previously formed α-MnBi phase. As a consequence, a significant increase of the α-MnBi phase during the second annealing step is obtained which is accompanied by a high magnetic coercivity.

**Materials and methods**

A schematic overview of the processing steps used in this study is shown in Fig. 1. Mn-Bi composites were processed using high-pressure torsion (HPT)[20]. Mn (99.95% - 325 mesh) and Bi powders (99.999% - 200 mesh) were blended and mixed at a ratio of 50 at.%. The powder blends were subsequently compressed (2 GPa pressure, room temperature, 0.6 rpm, ¼ rotation) into disk-shaped samples (8 mm diameter, ~0.7 mm thickness). Sufficient thickness for the subsequent deformation was ensured by filling the powder into a copper ring, which was fully removed after the HPT-compression step. To prevent oxidation, the above-described blending and compression process steps are carried out in Ar-atmosphere (Ar purity 99.999%). The compacted disks were at first HPT deformed at room temperatures to various amounts of deformation (1, 20 and 100 rotations) to investigate the effect of strain on the composite microstructure (1$^{st}$ step, Fig.1). The rotational speed and the applied pressure were 0.6 rpm and 2 GPa for all samples.

After HPT deformation, the samples were isothermally annealed in a magnetic field at 240°C for 4 h (2$^{nd}$ step, Fig.1). For this process step, a custom-made vacuum chamber and heating device is used, which is placed into the homogeneous field region provided by the conical pole pieces (diameter of 176 mm) of an electromagnet (Type B-E 30, Bruker). The electromagnet was operated at a constant magnetic field of 2T during heating. Two copper blocks containing cartridge heating elements (Ø10 x 50 mm, 100 W) are inside the vacuum chamber and heat the previously HPT deformed samples by a direct heat transfer through the copper block. A constant temperature is monitored by a thermocouple mounted next to the samples. The applied external magnetic field is recorded using a Hall-probe (Model 475 DSP, Lakeshore), which is placed outside of the vacuum chamber. The annealed disks were then HPT deformed at room temperature for 10 rotations with a rotational speed of 0.6rpm and an applied pressure of 2 GPa (3$^{rd}$ step, Fig.1). After the second HPT deformation step, the same annealing procedure as described above was applied (4$^{th}$ step, Fig.1). Basically, the two steps of HPT-deformation (step i) and magnetic field assisted annealing (step i+1) can also be repeated more than once.

Composite microstructures were analysed in tangential viewing direction using scanning electron microscopy (SEM, Zeiss Leo 1525, 20 kV acceleration voltage, probe current 200 nA) using the backscattered electron detector (60 µm aperture size, working distance 5-7 mm). For detailed microstructure analysis, electron backscatter diffraction (EBSD) measurements are performed with a Brukere-Flash$^{FS}$ detector. The X-ray diffraction (XRD)- measurements were obtained using a Bruker Phaser D2 device (Co-Kα radiation, 2θ=20-95°, exposure time 3.5 s per spot, 5000 steps, 1 mm aperture size, 0.6 mm collimator size). The radiation is applied to the axial surface normal of the HPT samples (parallel to magnetic field direction during annealing). To evaluate magnetic properties, the samples were measured at 300 K using a SQUID-magnetometer (Quantum Design MPMS-XL-7) operated with the manufacturer's software MPMSMultiVu Application (version 1.54) in tangential HPT direction.

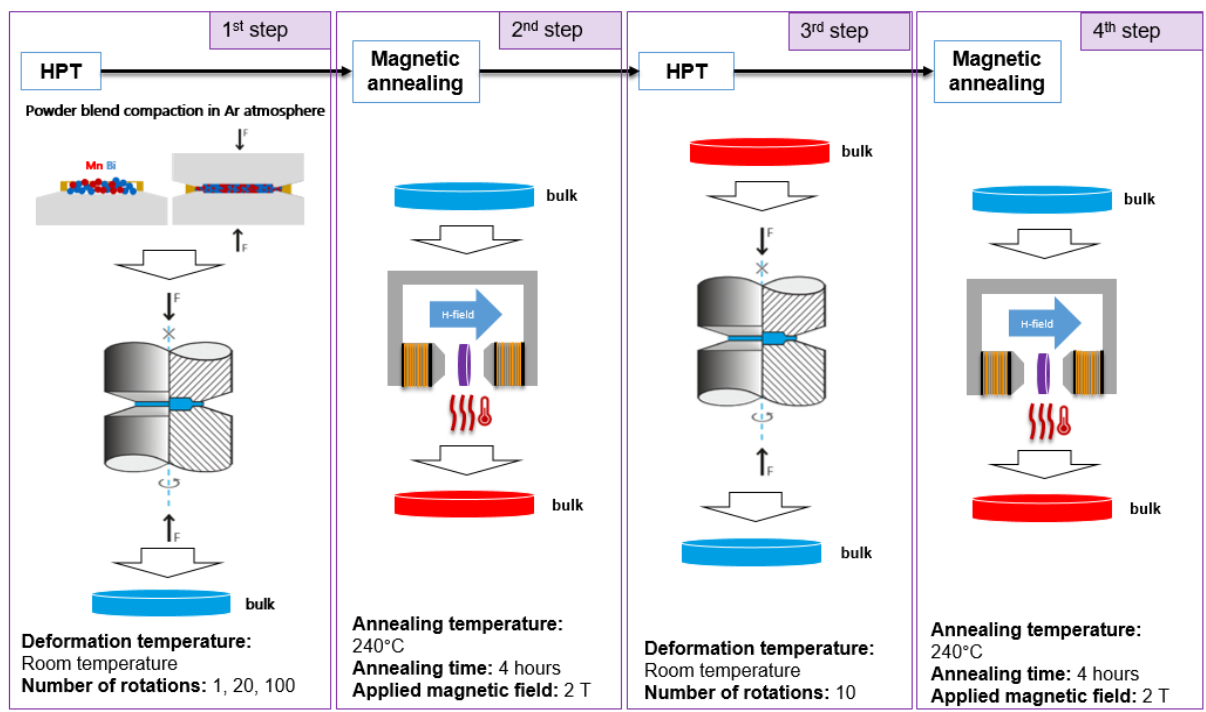

Fig. 1: Schematic overview of the processing steps. After each HPT step (1st, 3rd), the samples are annealed in a magnetic field (2nd, 4th).

**Results and discussion**

Fig.2 a displays representative microstructure of the Mn-Bi composites at a radius of 3 mm for different HPT rotations (N=1, 20, 100) with a low magnification to show the overall composite structure after deformation. Using backscattered electron signal in composition mode, lighter Mn appears dark and heavier Bi phase appears bright.

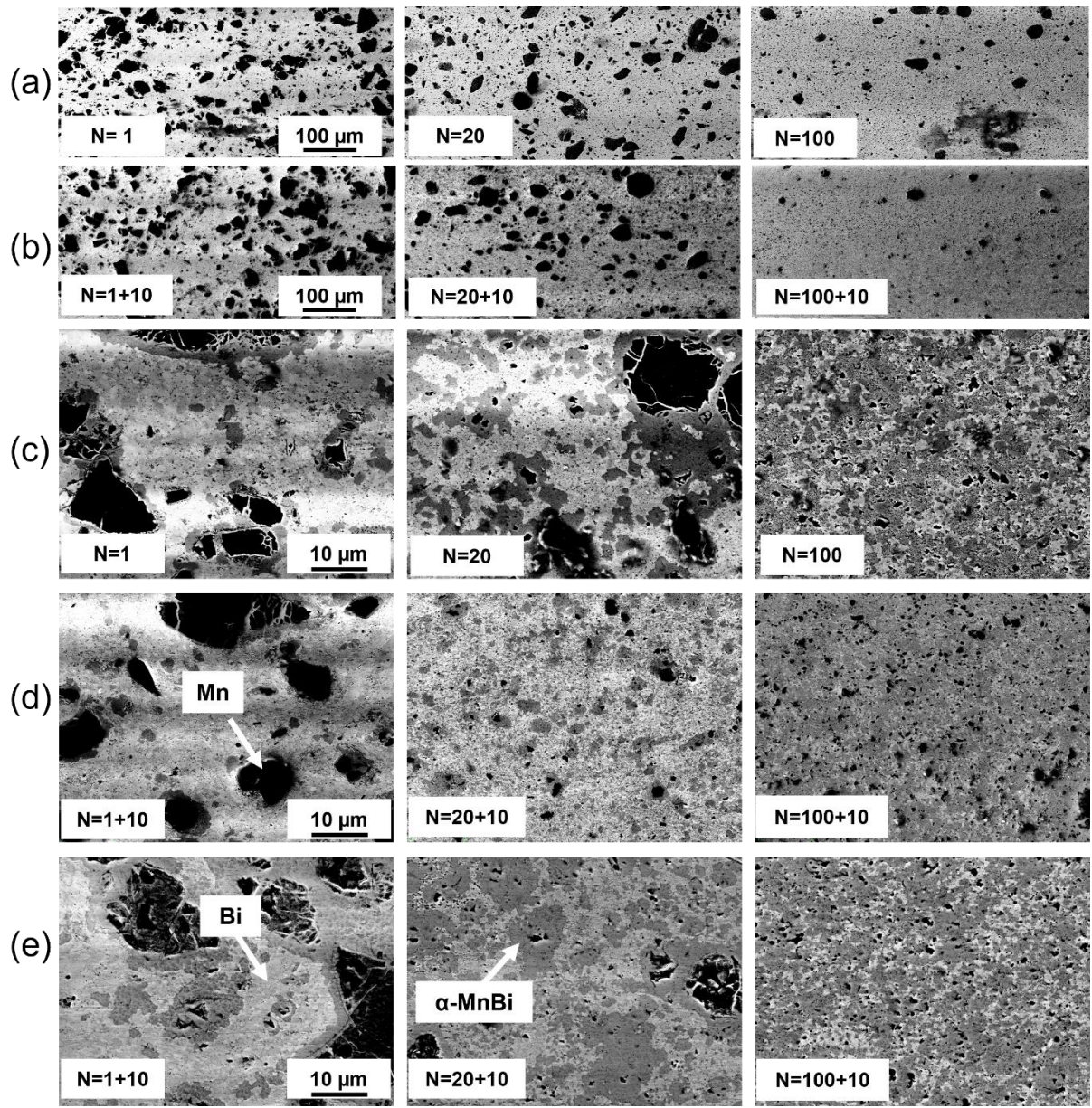

*Fig. 2: Backscattered electron images of the microstructure of the Mn-Bi composites at a radius of 3 mm after the 1$^{st}$ (a) and 3$^{rd}$ (b) processing step for samples deformed to a different number of rotations (N). Backscattered electron images of the microstructure of the Mn-Bi composites after the 2$^{nd}$ (c), 3$^{rd}$ (d) and 4$^{th}$ (e) processing step with higher magnification. Due to Z-contrast, the Mn phase appears darkest, Bi phase brightest and the α-MnBi phase has a medium "greyish" contrast. The scale bar in (a) to (e) applies to all micrographs in a row.*

For low two-stage deformation (N = 1-30), large Mn particles with a size up to ~50 µm are visible. Their size gets significantly refined with increasing applied strain. After 100 rotations, still a few large Mn particles are visible although the overall homogeneity of the Mn-Bi composite is improved. Fig.2 b displays representative microstructures of the Mn-Bi composites at a radius of 3 mm after the second HPT deformation (3$^{rd}$ step) with a low magnification. While no significant reduction in Mn particle size is evident for the samples deformed for 1 and 20 rotations in the first step, the Mn particle size in the composite with the highest number of rotations (N=100+10) decreases significantly. Only a few large Mn particles remain.

The structural refinement during HPT processing of a composite which contains a hard (Mn) and a soft phase (Bi) is characterized by strong plastic deformation in the soft phase. When the hard phase is present in a low volume fraction, a strong strain localization in the soft phase occurs, leading to a retardation of the refinement process of the hard phase [21]. The large residual Mn particles further affect the α-MnBi phase formation after annealing (3$^{rd}$ step). Fig.2c shows the microstructures after annealing of the Mn-Bi composites deformed for 1, 20, and 100 rotations. In these micrographs, the α-MnBi phase appears greyish. It is further visible that the α-MnBi phase preferentially forms around existing Mn particles. Refined Mn particles exhibit a higher surface area. Thus, the sample deformed for 100 rotations, shows an increased α-MnBi phase fraction after annealing as can be clearly seen in the respective micrographs (Fig.2c N=1 and N=100). The area fraction was further determined from the respective micrographs using the software program ImageJ [22], which confirms the higher α-MnBi phase fraction (14.8% for N=1 and 51.4% for N=100). In Fig. 2d, the microstructure of the same Mn-Bi composites after the additional HPT deformation (3$^{rd}$ step) with a higher magnification is shown. The second HPT deformation does not only affect overall homogeneity of the Mn and Bi phases in the composite, but also the phase fraction of the α-MnBi phase increases. After the last processing step (final magnetic annealing), a high amount of α-MnBi phase is visible in the micrographs obtained by SEM in this composite (Fig.2e). The area fraction from the micrographs was determined with ImageJ [22], which is 33.6% for N1+10, 42.3% for N20+10 and 58.9% for N100+10, respectively.

XRD analysis confirm, that the α-MnBi phase is formed for all annealed samples after the second processing step (Fig.3a-c). The radiation is applied to the axial surface normal of the HPT samples (parallel to magnetic field direction during annealing) and patterns are normalized with respect to the most intense Bi peak (2θ=31.6°). The Mn peaks (asterisk) are very low in intensity, however no α-MnBi phase (cross) is visible after the 1st processing step. After the 2nd step (annealing) the α-MnBi phase is identified, and its relative intensity is increased after the 4th step. No unidentified peaks or oxides are found.

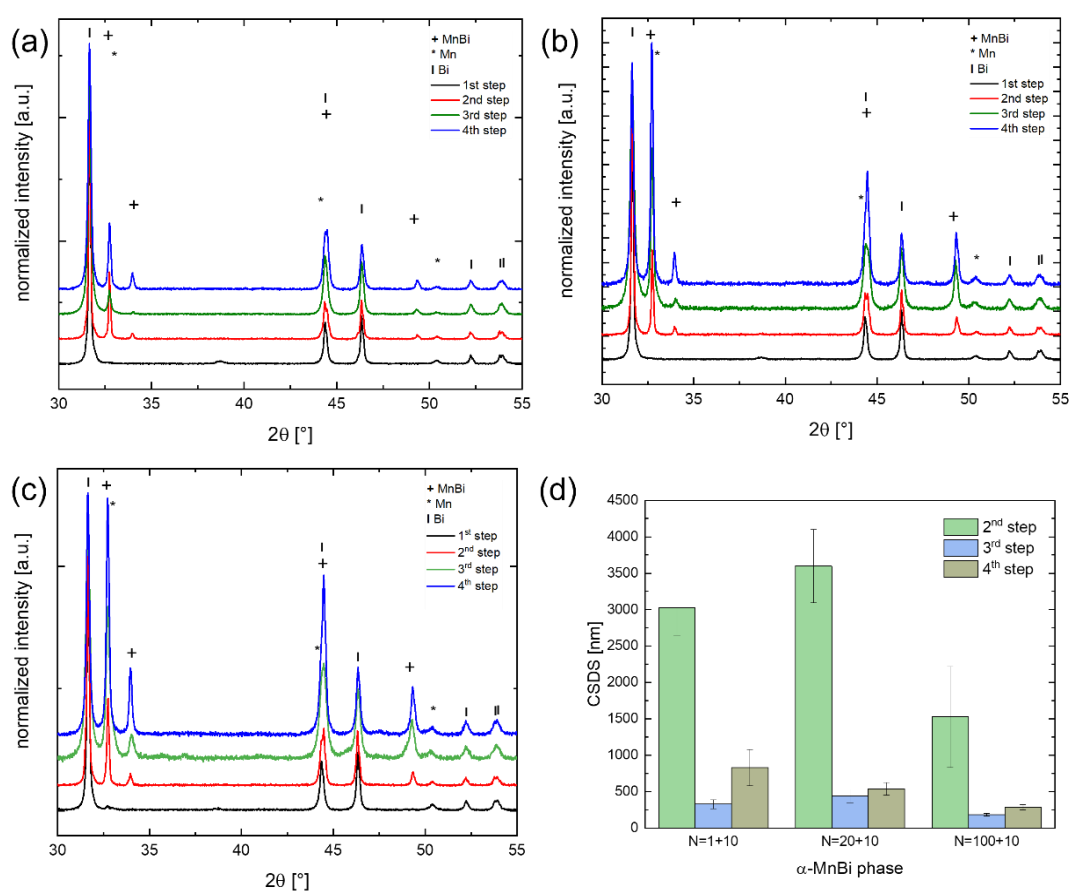

*Fig. 3: XRD pattern of all samples after the 1st, 2nd, 3rd and 4th processing step. (a) N=1 + 10 rotations, (b) N=20 +10 rotations, (c) N=100 + 10 rotations. References for expected peak positions are taken from the Crystallography Open Database (Mn: COD 9011108, Bi: COD 9008576, α-MnBi: COD 9008899). (d) ~~Crystallite size~~ CSDS of the α-MnBi phase after 2nd, 3rd and 4th processing step for all samples.*

To determine the coherent scattering domain size (CSDS) ~~crystallite size~~ of the α-MnBi phase, the measured XRD patterns were analysed by using the Rietveld refinement method using Profex (Fig.3d). The CSDS in the α-MnBi phase after the first annealing step is in the µm-range for all samples (c.f. Fig.3d). The CSDS of the α-MnBi phase is quite similar after 1 and 10 rotations. As the MnBi phase preferentially forms around existing Mn particles (c.f. Fig.2), the key to achieve a small grain size is to have a high number of refined Mn particles. The amount of the large Mn particles very similar at N=1 and N=20 (c.f. Fig.2). At N=100, however, the number of large Mn particles gets significantly less, which leads to a significantly reduced CSDS of the α-MnBi phase. The second HPT deformation then leads to structural refinement in the α-MnBi phase, where CSDS below 500 nm are obtained for all samples. However, the initial (2$^{nd}$ step, annealing), intermediate (3$^{rd}$ step, HPT) and final CSDS (4$^{th}$ step, annealing) are smallest in the samples deformed to the highest number of rotations (N=100+10).

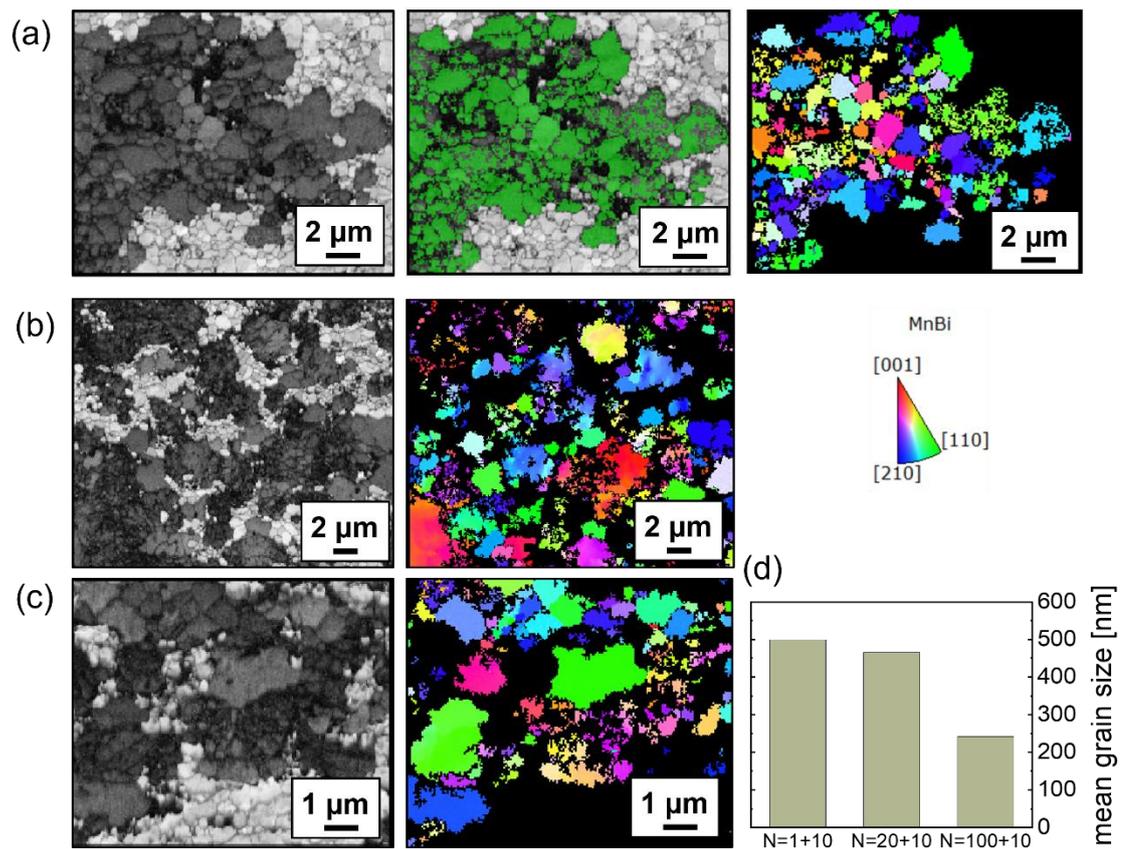

Fig. 4: EBSD image quality and orientation maps after the 4$^{th}$ processing step. (a) N=1 + 10 rotations, (b) N=20 +10 rotations, (c) N=100 + 10 rotations. Please note the different magnification. In (a), the image quality map is shown with a α-MnBi phase overlay (green). (d) Mean grain size of the α-MnBi phase after 4$^{th}$ processing step for all samples.

Fig. 4 shows representative image quality and orientation maps of the microstructures after the 4$^{th}$ step for (a) N=1 + 10 rotations, (b) N=20 +10 rotations, (c) N=100 + 10 rotations. In Fig.4a, the image quality map is further overlayed by a green phase overlay of the α-MnBi phase, which illustrates that the α-MnBi phase is the dark grey area visible in the image quality maps. From the image quality maps it can be seen that the α-MnBi phase consists of grains with different sizes for all processing conditions, which results in a broad grain size distribution. In Fig.4d, the mean grain size of the α-MnBi phase obtained from several EBSD scans after 4$^{th}$ processing step is shown. As the spatial resolution is inherently limited by the pattern source volume, EBSD is insufficient to accurately detect truly nanostructured grains (with sizes below 100 nm) as can be seen by comparison of the image quality and corresponding orientation maps. However, the same tendency of decreasing grain sizes with increasing number of rotations as observed by XRD measurements is obtained (Fig.3d).

Fig.5 shows the hysteresis loops recorded between an applied field of ±60 kOe in tangential HPT-disk direction of all samples after the 4$^{th}$ and final processing step and the obtained magnetic properties are summarized in Table 1. Values for the saturation magnetization ($M_s$) are acquired by linearly extrapolating the magnetization against $H^{-1}$ and determining the intercept at $M(H^{-1}) = 0$ [23,24]. For the sample deformed to the highest number of rotations (N=100+10), a $M_s$ of 55.1 emu/g is found. The coercivity $H_c$ increases from 1.4 kOe (N=1+10), 2.2 kOe (N=20+10) to 3.1 kOe (N=100+10).

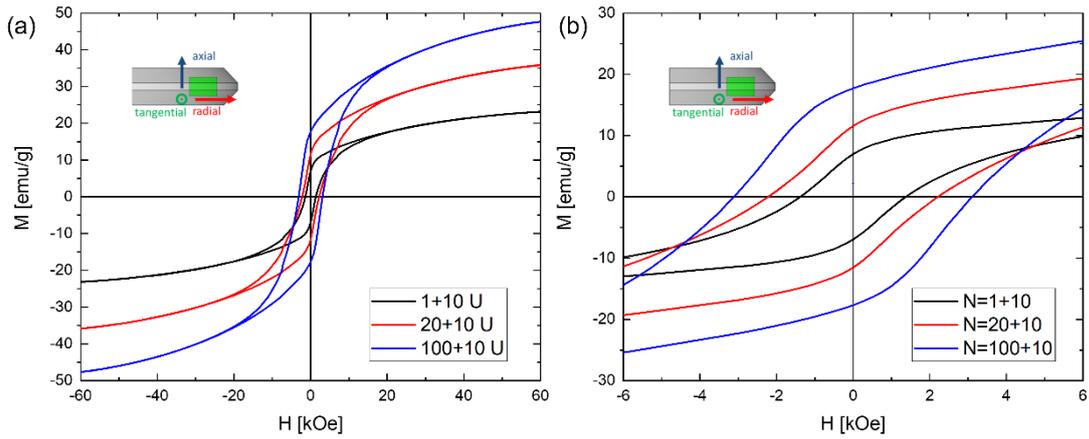

*Fig. 5: (a) SQUID magnetometry hysteresis loops (±60 kOe) and (b) magnified section for the samples after the 4th processing step measured in tangential HPT direction as shown in the insets. The green cubes in the insets represent the SQUID specimen cut from the HPT sample.*

In Table 1, the α-MnBi content in weight percent is further given. It is calculated using a linear approximation with the measured saturation magnetization (Ms) and with the theoretical Ms of 79 emu/g of the α-MnBi phase, neglecting the paramagnetic Mn and the diamagnetic Bi phases [25]. The α-MnBi area fraction obtained from SEM micrographs using the software program ImageJ [22] increases from 33.6% (N=1+10), 42.3% (N=20+10) to 58.9% (N=100+10).

*Table 1: Magnetic properties taken from hysteresis loops presented in Fig. 5. The determined coercivity $H_c$, saturation magnetization $M_s$, remnant magnetization $M_r$, maximum energy product ($BH_{max}$) and calculated and from SEM micrographs determined α-MnBi phase fraction are listed.*

| Rotations (N) | $H_c$ (kOe) | $M_r$ (emu/g) | $M_s$ (emu/g) | $BH_{max}$ (kJ/m³) | α-MnBi (wt%), calculated | α-MnBi (%), SEM |
|---|---|---|---|---|---|---|
| 1+10 | 1,4 | 6,9 | 26,9 | 0,6 | 34,1 | 33,6 |
| 20+10 | 2,2 | 11,5 | 41,5 | 1,6 | 52,6 | 42,3 |
| 100+10 | 3,1 | 17,7 | 55,1 | 4,4 | 69,7 | 58,9 |

It is known from several studies that the application of a magnetic field during annealing has a beneficial influence on the α-MnBi phase formation and the resulting α-MnBi phase enhancement as it might affect the formation enthalpy of the α-MnBi phase, the activation energy of the reaction and the atomic diffusion between Mn and Bi [9,10,12,26–28]. To achieve a large amount of α-MnBi phase during annealing, an increase of the number of Mn- and Bi-interfaces due to the grain size-reduction supports the grain boundary diffusion of Mn and Bi [12]. Our experiments also clearly show that the highest content of α-MnBi phase is obtained in the sample after the utmost number of HPT deformation turns (N=100+10) as this sample exhibits the highest number of submicron-Mn particles with an increased area between Mn and Bi interfaces. From SEM measurements, 58.9 % α-MnBi and 41.1 % residual Bi and Mn phase is present.

On the other hand, a high defect density is beneficial to obtain a large amount of α-MnBi phase during annealing. This high defect density in severe-plastically deformed materials accelerates diffusion processes, which facilitate α-MnBi phase formation and its subsequent growth during annealing [30,31].

The second deformation step has also a beneficial effect in these samples as confirmed by XRD measurements and comparison of the peak intensities of the α-MnBi phase (c.f. Fig. 3a-c). It is important to note that this second deformation step and thus the SPD of the α-MnBi phase is only possible, because after the first annealing step, a large volume of bulk α-MnBi material is already present. During the second deformation step, phase and grain sizes of the remaining Mn and Bi phase as well as the previously formed α-MnBi phase are refined leading to new Mn-Bi interfaces and enable further α-MnBi phase formation during the second annealing step. During the first annealing step, recovery of the severe plastically deformed microstructure is expected. Repeated HPT deformation increases the defect density again to accelerate diffusion processes and ease α-MnBi phase formation and subsequent phase growth during the final annealing. In principle, the two steps of HPT-deformation (step i) and magnetic field assisted annealing (step i+1) can also be repeated more than once with the possibility of achieving a nearly single-phase α-MnBi after a sufficient number of repetitions. Another option is to make the SPD process continuous by carrying out the magnetic field assisted annealing directly in the HPT device. In preliminary experiments, a magnetic field of 0.4 T has already been applied. The use of a different SPD process can also be considered, for example, accumulative roll bonding in combination with magnetic annealing. It is hypothesized that the high amount of phase interfaces generated during accumulative roll bonding and the pre-textured Mn and Bi crystals provide favourable nucleation sites for magnetic-field-annealing induced growth of textured α-MnBi grains leading to sheet materials with high amounts of α-MnBi phase, ideally close to 100%.

The coercivity values obtained in this study are higher in comparison to values of bulk α-MnBi magnets manufactured from Mn and Bi in [9,22], which might be related to magnetic hardening by grain refinement [32,33]. Although >90% α-MnBi volume fraction is obtained in [9], coercivity of pure bulk α-MnBi magnets without additional alloying elements is lower ($H_c$ =0.6k Oe). Higher coercivity values of 2.2k Oe are obtained in [27], but a high field of 15 T was applied in the solid-state reaction sintering process. Bulk α-MnBi magnets without addition of alloying elements fabricated in [15,16] achieve similar magnetic properties as described here. In [13,14,17], however, significant better magnetic properties compared to our study are achieved.

Although during the last annealing step, grain growth of the α-MnBi phase is observed, the CSDS and mean grain size of the α-MnBi phase are below 1 µm (833 nm (N=1+10), 540 nm (N=20+10) and 286 nm (N=100+10) and 499 nm (N=1+10), 466 nm (N=20+10) and 242 nm (N=100+10) as determined by XRD and EBSD, respectively). For the sample deformed to the highest number of rotations (N=100+10), the value is furthermore close to the single domain size of α-MnBi (about 250 nm [36]) for which a large $H_c$ is expected. Bulk α-MnBi magnets with higher $H_c$ have been manufactured from melt-spun alloys modified with Mg and Sb [9]. Besides further repetition of HPT-deformation and magnetic field assisted annealing steps to achieve nearly single-phase α-MnBi, future work will also concentrate on modification of the initial composition of the HPT-deformed composites as the process allows a free selection of starting composition and the addition of other elements.

## Conclusions

The new processing route proposed in this study offers an easy way to produce non-rare earth bulk magnetic material with a high α-MnBi phase amount avoiding a large number of different processing steps. A high amount of α-MnBi phase was obtained for the multistep HPT processed Mn-Bi composites in combination with magnetic field assisted annealing. A second deformation step promoted phase and grain refinement of the previously formed α-MnBi phase in combination with homogeneous nanocomposite structures. Thus, final magnetic field assisted annealing leads to α-MnBi phase amount as high as 70 wt.% with a coercivity of 3.1 kOe. In principle, the process of deformation and annealing treatment presented in this study can be repeated more often and for longer isothermal annealing times. Additional HPT deformation and annealing steps are expected to allow successful synthesis of α-MnBi based composites with an even higher amount of α-MnBi phase. Therefore, multistep HPT processing with up-scaling potential [37,38] might become attractive for future industrial applications.

## Acknowledgements

This project has received funding from the European Research Council (ERC) under the European Union's Horizon 2020 research and innovation programme (Grant No. 757333 and 101069203).

## Disclosure statement

The authors report there are no competing interests to declare.

## Data availability statement

The datasets used and/or analysed during the current study are available from the corresponding author on reasonable request.